\documentclass{ws-ijmpe}
\begin{document}

\markboth{F. Iazzi, R. Introzzi, A. Lavagno, D. Pigato, M.H. Younis}{Strangeness production at finite temperature and baryon density in an effective relativistic mean field model}

\catchline{}{}{}{}{}

\title{Strangeness production at finite temperature and baryon
density in an effective relativistic mean field model}

\author{F. Iazzi, R. Introzzi, A. Lavagno, D. Pigato, M.H. Younis}
\address{Dipartimento di Fisica, Politecnico di Torino, I-10129 Torino, Italy\\
INFN, Sezione di Torino, I-10126 Torino, Italy}

\maketitle

\begin{history}
\received{(received date)}
\revised{(revised date)}
\end{history}

\begin{abstract}
We study the strangeness production in hot and dense nuclear medium, by requiring the conservation of the baryon density, electric charge fraction and zero net strangeness. The hadronic equation of state is investigated by means of an effective relativistic mean field model, with the inclusion of the full octet of baryons and kaon mesons. Kaons are considered taking into account of an effective chemical potential depending on the self-consistent interaction between baryons. The obtained results are compared with a minimal coupling scheme, calculated for different values of the anti-kaon optical potential and with non-interacting kaon particles. In this context, we also consider the possible onset of the kaon condensation for a wide range of temperatures and baryon densities.
\end{abstract}

\section{Introduction}

In the last years there has been a growing interest in the study of the kaons and anti-kaons proprieties in finite nuclear matter as well as in compact stars. Many studies have shown the possibility that these particles may condensate in neutron stars, whereas the situation seems to be more uncertain in the physical conditions reachable in relativistic heavy ion collision  \cite{zakout,pal,thorsson,kaplan1,banik,hartnack,plb_quarati}. This uncertainty is mainly due to the difficulty of finding out from the experimental results an estimation of the real part of the anti-kaon optical potential at the nuclear saturation density. A stronger attractive potential depth should in fact favor the formation of the condensation, but, on the other hand, a stiffer or softer equation of state (EOS) should help or not the condensation itself. At this regard, the analysis of kaonic atom data leads to the real part of anti-kaon optical potential close to $U_{K^-} = -180 \pm 20$ MeV at the saturation nuclear matter density \cite{banik,fried1,batty,fried2,glena,pons}. Contrariwise, chiral-models and coupled-channel G-matrix theory, seem to suggest a strength of the optical potential close to $U_{K^-} =-(50 \div 80)$ MeV \cite{tolos,lutz,lutz2}.
These uncertainties in the estimation of the anti-kaon potential depth, imply some difficulties in the calculation of the effective kaon mass in-medium. Moreover, different mean field models predict negative or imaginary effective kaon mass at sufficiently large values of the $\sigma$-meson field, responsible of the medium range attraction \cite{glena,pons,knorren}.
At this regard, the future CBM (Compressed Baryonic Matter) experiment of FAIR (Facility
of Antiproton and Ion Research) at GSI Darmstadt, will be of great importance to create compressed baryonic matter with a high net baryon density and finite temperature \cite{senger,henning,arsene,bravina} and make possible an accurate analysis of the proprieties of kaons and, more in general, of the strangeness production at high baryon density.

In this work, we are going to investigate the strangeness production at finite isospin density and zero net strangeness, using two different approaches for what concern the kaon degrees of freedom. The first one consists in a relativistic mean field model, with the introduction of an effective meson chemical potential, obtained through a self consistent interaction between baryons \cite{andrea}. The second one is based on a minimal coupling scheme, where kaons are directly coupled with the scalar and vectors meson fields ($\sigma$, $\omega$ and $\rho$) through coupling constants related to the anti-kaon optical potential \cite{banik,glena,pons,cavagnoli}.

We focalize our analysis in the range of temperature and baryon density reachable in high energy heavy ion collisions with particular attention to the physical conditions relevant for the future CBM experiment at FAIR. At this stage, we do not take in consideration the possible formation of a mixed phase or deconfinement phase transition to a quark-gluon plasma. We point our attention in exploring the  strangeness production of a hadronic system and the possible onset of the kaon condensation for the two aforementioned models. In this context, other strangeless mesons are not considered in our analysis, because they do not sensibly affect the strangeness production but they contribute mainly to the total pressure and energy density.

The paper is organized as follow. In the Sect. $2$, we introduce the hadronic equation of state and the two frameworks (minimal coupling scheme and effective relativistic mean field model) with which we analyze the kaons behavior in the nuclear medium, for various values of the anti-kaon optical potential. In Sect. $3$, we report and discuss our major results obtained in this comparative study, and finally in Sect. $4$ we summarize our conclusions.

\section{Hadronic equation of state}
The relativistic mean field model (RMF), first introduced by Walecka and Boguta-Bodmer \cite{walecka,boguta}, is widely successful used for describing the
properties of finite nuclei as well as hot and dense nuclear matter
\cite{serot,glen0,glen1,bonanno,alberico,plb,prd2004,epja2011}. In this context, the total baryon Lagrangian density can be
written as
\begin{equation}
{\cal L}_B={\cal L}_{\rm octet}+{\cal L}_K \, ,
\end{equation}
where ${\cal L}_{\rm octet}$ stands for the full octet of baryons
($p$, $n$, $\Lambda$, $\Sigma^+$, $\Sigma^0$, $\Sigma^-$, $\Xi^0$,
$\Xi^-$) and ${\cal L}_K$ corresponds to the kaon meson degrees of freedom.

The quantum hadrodynamics model for the full octet of baryons ($J^P=1/2^+$) was
originally studied by Glendenning with the following standard
Lagrangian \cite{glen0}
\begin{eqnarray}\label{lagrangian}
{\cal L}_{\rm octet} &=&
\sum_i\bar{\psi}_i\,[i\,\gamma_{\mu}\,\partial^{\mu}-(m_i- g_{\sigma
i}\,\sigma) -g_{\omega
i}\,\gamma_\mu\,\omega^{\mu} -g_{\rho i}\,\gamma_{\mu}\,\vec{\tau}
\cdot \vec{\rho}^{\;\mu}]\,\psi_i \nonumber\\
&&+\,\frac{1}{2}(\partial_{\mu}\sigma\partial^{\mu}\sigma-m_{\sigma}^2\sigma^2)
-U(\sigma)+\frac{1}{2}\,m^2_{\omega}\,\omega_{\mu}\omega^{\mu} +\,\frac{1}{2}\,m^2_{\rho}\,\vec{\rho}_{\mu}\cdot\vec{\rho}^{\;\mu} \nonumber\\
&&-\frac{1}{4}F_{\mu\nu}F^{\mu\nu} -\frac{1}{4}\vec{G}_{\mu\nu}\vec{G}^{\mu\nu}\, ,
\end{eqnarray}
where the sum runs over all the baryons octet, $m_i$ is the vacuum
baryon mass of index $i$, $\vec{\tau}=2\vec{t}$ denotes the isospin
operator which acts on the baryon and $U(\sigma)$ is the nonlinear
self-interaction potential of $\sigma$ meson
\begin{eqnarray}
U(\sigma)=\frac{1}{3}a\,(g_{\sigma
N}\,\sigma)^{3}+\frac{1}{4}\,b\,(g_{\sigma N}\,\sigma^{4})\, ,
\end{eqnarray}
introduced by Boguta and Bodmer to achieve a reasonable
compressibility for equilibrium normal nuclear matter \cite{boguta,serot,glen1}.

The meson-nucleon couplings constant ($g_{\sigma N}$, $g_{\omega N}$ and $g_{\rho N}$)
have been fixed to the parameters set marked as GM3 of Ref.\cite{glen1}.
The implementation of the model with the inclusion of the hyperons
requires the determination of the corresponding meson-hyperon
coupling constant, that have to be fitted to the hypernuclear proprieties.
This can be done by fixing the scalar $\sigma$-meson hyperon coupling
constants to the potential depth of the corresponding hyperon in
saturated nuclear matter ($U_{\Lambda}^N=-28$ MeV, $U_{\Sigma}^N= 30$ MeV,
$U_{\Xi}^N=-18$ MeV) \cite{gmuca,mllener,schaffner}.
The obtained ratios $x_{\sigma Y}=g_{\sigma Y}/g_{\sigma N}$ for the GM3 parameter set are:
$x_{\sigma \Lambda}=0.606$, $x_{\sigma \Sigma}=0.328$ and $x_{\sigma \Xi}=0.322$.
Furthermore, following
Ref.s \cite{schaffner_prl1993,schaffner_annphys1994},
the SU(6) simple quark model can be used to obtain the relations
\begin{eqnarray}
\frac{1}{3}g_{\omega N}=\frac{1}{2}g_{\omega
\Lambda}=\frac{1}{2}g_{\omega \Sigma}=g_{\omega\Xi} \, , \nonumber\\
g_{\rho N}=\frac{1}{2}g_{\rho\Sigma}=g_{\rho\Xi}\,, \ \ \ g_{\rho
\Lambda}=0 \, .
\end{eqnarray}

Because we are going to describe finite temperature and density
nuclear matter with respect to strong interaction, we have to
require the conservation of three "charges": baryon
number, electric charge and strangeness number. As a consequence, we have to consider three independent chemical potentials: $\mu_B$,
$\mu_C$ and $\mu_S$, respectively, the baryon, the electric (isospin) charge
and the strangeness chemical potentials of the system \cite{ditoro,jpg2010}. Hence,
the chemical potential of particle of index $i$ can be written as
\begin{equation}
\mu_i=b_i\, \mu_B+c_i\,\mu_C+s_i\,\mu_S \, ,
\label{mu}
\end{equation}
where $b_i$, $c_i$ and $s_i$ are, respectively, the baryon, the
electric charge and the strangeness quantum numbers of the $i$-th
hadronic species.

Kaons degrees of freedom are treated here in two distinct approaches: the first one, well known in literature, considers the interaction between kaons and nucleons by means of a direct minimal coupling scheme with the meson fields ($\sigma$, $\omega$, $\rho$)
\cite{banik,glena,pons,cavagnoli}, the second one considers the meson degrees of freedom as a quasi-ideal Bose gas, with an effective chemical potentials $\mu^*$ depending
on the self-consistent interaction between baryons \cite{andrea}.

In the first case, we can write the kaon Lagrangian density in terms of minimal coupling
scheme, as follow \cite{banik,glena,pons,cavagnoli}
\begin{eqnarray}\label{eq:lagrangianaK}
{\mathcal L}_{K}= D_{\mu}^*\Phi^* D^{\mu}\Phi -m_{K}^{*2}\Phi^*\Phi\, ,
\end{eqnarray}
where $D_{\mu}=\partial_{\mu} +ig_{\omega K}\omega_{\mu} +ig_{\rho
K}\tau_{3 K}\rho_{\mu}$ is the covariant derivative of the meson field,
$m^*_{K}=m_K -g_{\sigma K}\sigma$ is the effective kaon mass and
$\tau_{3 K}$ is the third component of the isospin operator.

The kaon-meson vector coupling constant are obtained in the usual way,
from the quark model and isospin counting rules, setting
$g_{\omega K}= g_{\omega N}/3$ and $g_{\rho K}=g_{\rho N}$.
Whereas the scalar $g_{\sigma K}$ coupling constant is determined from the
study of the real part of the anti-kaon optical potential, at the
saturation nuclear density, in a symmetric nuclear matter, by setting
$U_{K^-}= -g_{\sigma K}\sigma - g_{\omega K}\omega$.

In this investigation we set the
anti-kaon optical potential equal to $U_{K^-}=-50$ MeV, $-100$ MeV and $-160$ MeV,
based on recent theoretical calculations and experimental measurements
\cite{banik,fried1,fried2,glena,pons,tolos,lutz,lutz2,gla}. For these values of the anti-kaon potential depth, we obtain the following kaon optical potentials ($U_{K^+}=-g_{\sigma K}\sigma + g_{\omega K}\omega$, where the sign $+$ in the $\omega$-field, is due to the G-parity): $U_{K^+}= 47$ MeV, $-3$ MeV and $-63$ MeV, respectively for $U_{K^-}=-50$ MeV, $-100$ MeV and $-160$ MeV, at the saturation nuclear density. In this approach we neglect the contribution of the neutral kaons ($K^0$ and $\overline{K}^0$) because of the large uncertainty in literature on their coupling constants with the meson fields.

The field equation are obtained in the usual way, from the minimization of the thermodynamic potential $\Omega_H=\Omega_B +\Omega_K $, with respect to the $\sigma$, $\omega$ and $\rho$ meson fields
\begin{eqnarray}
&& m^2_{\sigma}\sigma = -\frac{\partial U}{\partial \sigma} +\sum_{i} g_{\sigma i}\rho^S_i  + g_{\sigma K} \, (\rho^S_K + \rho^c_K)\, ,\\
&& m^2_{\omega}\omega = \sum_{i} g_{\omega i}\rho_i +  g_{\omega K} \, \rho_K \, ,\\
&& m^2_{\rho}\rho =  \sum_{i} g_{\rho i}\tau_{3i}\rho_i + g_{\rho K} \, (\rho_{K^+}-\rho_{K^-}) \, ,
\label{eq:eqcampi}
\end{eqnarray}
where $\sigma=\langle\sigma\rangle$, $\omega=\langle\omega^0\rangle$
and $\rho=\langle\rho^0_3\rangle$ are the nonvanishing expectation
values of mesons fields, $\rho_i^B$ and $\rho_i^S$ are respectively the
vector and baryon scalar density of the baryon particle
of index $i$:
\begin{eqnarray}
&&\rho_i^B=\gamma_i \int\frac{{d}^3k}{(2\pi)^3}\;[n_i(k)-\overline{n}_i(k)] \, , \\
&&\rho_i^S=\gamma_i \int\frac{{d}^3k}{(2\pi)^3}\;\frac{m_i^*}{E_i^*}\;
[n_i(k)+\overline{n}_i(k)]\, ,
\end{eqnarray}
where $\gamma_i=2J_i+1$ is the degeneracy spin factor of the
$i$-th baryon ($\gamma_{\rm octet}=2$), $n_i(k)$ and $\overline{n}_i(k)$ are the
fermion particle and antiparticle distributions function:
\begin{eqnarray}
n_i(k)&=&\frac{1}{\exp[(E^*_{i}(k)-\mu^*_i)/T]-1}\, ,\\
\overline{n}_i(k)&=&\frac{1}{\exp[(E^*_{i}(k)+\mu^*_i)/T]-1}\, .
\end{eqnarray}
The baryon effective energy $E_i^*$ is defined as $E_i^*(k)=\sqrt{k^2+{{m_i}^*}^2}$, with $m^*_i=m_i-g_{\sigma i}\sigma$ is the effective baryon mass and the effective chemical potentials $\mu_i^*$ are defined as
\begin{equation}
\mu_i^*=\mu_i-g_{\omega i}\,\omega-g_{\rho i}\,\tau_{3 i}\,\rho\, .
\label{mueff}
\end{equation}
The kaons vector and scalar density ($\rho_K$ and $\rho^S_K$),
are given respectively by \cite{cavagnoli}
\begin{eqnarray}
&&\rho_{K}= 2\xi^2(\mu_{K} - X_0) + \int\frac{{d}^3p}{(2\pi)^3}\;[n_K(p)- \overline{n}_K(p)] \, ,\label{eq:rhok} \\
&&\rho^S_K= \int\frac{{d}^3p}{(2\pi)^3}\,\frac{m^*_K} {\sqrt{p^2+m^{*2}_K}}\; [n_K(p) +\overline{n}_K(p)] \, ,  \label{eq:rhoks}
\end{eqnarray}
where we set $\mu_K=\mu_{K^+}$, $X_{0}= g_{\omega K}\omega_0 +g_{\rho K} \rho_0$, $\xi$ is the order parameter, obtained from the minimization of the thermodynamical
potential $\Omega_K$  (naturally for $\xi=0$ we have no condensation). Whereas, the kaon and anti-kaon distribution function are given by
\begin{eqnarray}
n_K(p)&=&\frac{1}{\exp[(\omega^{+}(p)-\mu_{K})/T]-1}\, ,\\
\overline{n}_K(p)&=&\frac{1}{\exp[(\omega^{-}(p) +\mu_{K})/T]-1}\, ,
\end{eqnarray}
where $\omega^{\pm}(p)=\sqrt{p^2+ m^*_K} \,  \pm g_{\omega K}\omega_0 +g_{\rho K}\tau_{3 K}\rho_{0}$ stays respectively for kaons and anti-kaons effective energy and the sign ($\pm g_{\omega K}\omega_0$) is due to the G-parity.

The kaon vector density in Eq.(\ref{eq:rhok}), can be viewed therefore as the sum of a "condensate" and a "thermal" contribution:
\begin{eqnarray}
\rho_K&=& \rho^c_K(\xi) + \rho^T_K(T)\, . \label{eq:rhok2}
\end{eqnarray}

The total pressure and energy density is given as usual by $P=P_B+P_K$ and
$\epsilon=\epsilon_B+\epsilon_K$, where
\begin{eqnarray}
P_B&=&\frac{1}{3}\sum_i \,\gamma_i\,\int \frac{{d}^3k}{(2\pi)^3}
\;\frac{k^2}{E_{i}^*(k)}\; [n_i(k)+\overline{n}_i(k)]
-\frac{1}{2}\,m_\sigma^2\,\sigma^2 - U(\sigma)  \nonumber\\
&+&\frac{1}{2}\,m_\omega^2\,\omega^2+\frac{1}{2}\,m_{\rho}^2\,\rho^2\, ,\\
\epsilon_B&=&\sum_i \,\gamma_i\,\int \frac{{
d}^3k}{(2\pi)^3}\;E_{i}^*(k)\; [n_i(k)+\overline{n}_i(k)]
+\frac{1}{2}\,m_\sigma^2\,\sigma^2+U(\sigma)\nonumber \\
&+&\frac{1}{2}\,m_\omega^2\,\omega^2 +\frac{1}{2}\,m_{\rho}^2 \,\rho^2\, ,
\end{eqnarray}
and for the kaons one gets
\begin{eqnarray}
P_K&=&\xi^2[(\mu_{K}-X_0)^2-m^{*2}_{K}] -T \int
\frac{{d}^3p}{(2\pi)^3} \{ \ln[1-e^{-\beta(\omega^+ -\mu_{K})}]\nonumber\\
&+& \ln[1-e^{-\beta(\omega^- +\mu_{K})}] \}\, , \label{eq:eosPK}\\
\epsilon_K&=&\xi^2[m^{*2}_K +\mu^2_{K} -X^2_0] + \int
\frac{{d}^3p}{(2\pi)^3} [\omega^+n_K(p) + \omega^-\overline{n}_K(p)]\, .
\label{eq:eosEK}
\end{eqnarray}
The condition for the onset of the kaon condensation, is then given by \cite{cavagnoli}
\begin{eqnarray}
\xi [\mu_K- \omega^+(0)][\mu_K +\omega^-(0)]=0 \, ,
\end{eqnarray}
therefore, for a s-wave condensation at $p=0$, we obtain, respectively, $\mu_{K^+}=\omega^+$ for $K^+$ and $\mu_{K^-}=\omega^-$ for $K^-$ (naturally when the condensate is not present, $\xi=0$).

In the second approach, we use an alternative formulation for what concern kaon degrees of freedom, based on the self-consistent interaction between baryons \cite{andrea}. In this scheme mesons are treated as a quasi ideal Bose gas with an effective chemical potential $\mu^*$, obtained from the bare one given in Eq.(\ref{mu}) and subsequently expressed in terms of the corresponding effective baryon chemical potentials, respecting the strong
interaction. More explicitly, the kaons and anti-kaon effective chemical potential can be written as
\begin{eqnarray}
&\mu^*_{K^+} = (\mu^*_p -\mu^*_{\Lambda}) = \mu_p -\mu_{\Lambda} - (1- x_{\omega \Lambda})g_{\omega N}\omega -\frac{1}{2}g_{\rho N}\rho\ ,
\label{mueffkp}\\
&\mu^*_{K^-} = (\mu^*_{\Lambda}-\mu^*_p) =\mu_{\Lambda} -\mu_p + (1- x_{\omega \Lambda})g_{\omega N}\omega +\frac{1}{2}g_{\rho N}\rho\, , \label{mueffkm}
\end{eqnarray}
where $x_{\omega \Lambda}=g_{\omega \Lambda}/g_{\omega N}$.

Thus, the effective meson chemical potentials are coupled with the meson fields related to the
interaction between baryons. This assumption represents a crucial feature in the EOS at finite
density and temperature and can be seen somehow in analogy with the hadron resonance gas within the excluded-volume approximation. There the hadronic system is still regarded as an ideal gas but in the volume reduced by the volume occupied by constituents (usually assumed as a phenomenological model parameter), here we have a (quasi-free) meson gas with an effective chemical potential that contains the self-consistent interaction of the meson fields.

Following the above scheme, we can evaluate the pressure $P_K$ and the energy density $\epsilon_K$ as a quasi-particle kaon gas considering the above effective meson potentials depending on the self-consistent interaction between baryons:
\begin{eqnarray}
P_K&=&\frac{1}{3} \,\int \frac{{d}^3p}{(2\pi)^3}\;\frac{p^2}{E_{K}(p)}\; [n_K(p)+\overline{n}_K(p)]\, ,\\
\epsilon_K&=&\int \frac{{d}^3p}{(2\pi)^3}\;E_{K}(p)\; [n_K(p)+\overline{n}_K(p)]\, ,\\
\rho_{K}&=&\int \frac{{d}^3p}{(2\pi)^3}\; [n_K(p)-\overline{n}_K(p)]\, ,
\end{eqnarray}
where $n_K(p)$ is the meson particle distribution function, given by
\begin{eqnarray}
n_K(p)&=&\frac{1}{\exp[(E_K(p)-\mu^*_K)/T]-1}\, ,
\end{eqnarray}
where $E_K(p)=\sqrt{p^2+m^2_K}$ and the corresponding antiparticle distribution $\overline{n}_K$ is obtained by substituting $\mu^*_K \rightarrow -\mu^*_K$. Finally, as before, the total pressure and the total energy density are simply the sum of the mesonic and baryonic components.

\section{Results and discussion}

Let us start our numerical investigation by studying in Fig. 1 the kaon to anti-kaon ratio
$K^+/K^-$ as a function of baryon density for different values of the anti-kaon optical potential. As expected, the ratio results to be very sensitive to the choices of the anti-kaon potential depth. At fixed temperature and by increasing the baryon density, we observe a continuous growing in the $K^+/K^-$ ratio for moderate values of $U_{K^-}$, whereas in presence of a strong attractive potential, the ratio decreases after reaching the nuclear saturation density, mainly due to the strong reduction of the kaon effective mass.
\begin{figure}[ht]
\centerline{\psfig{file=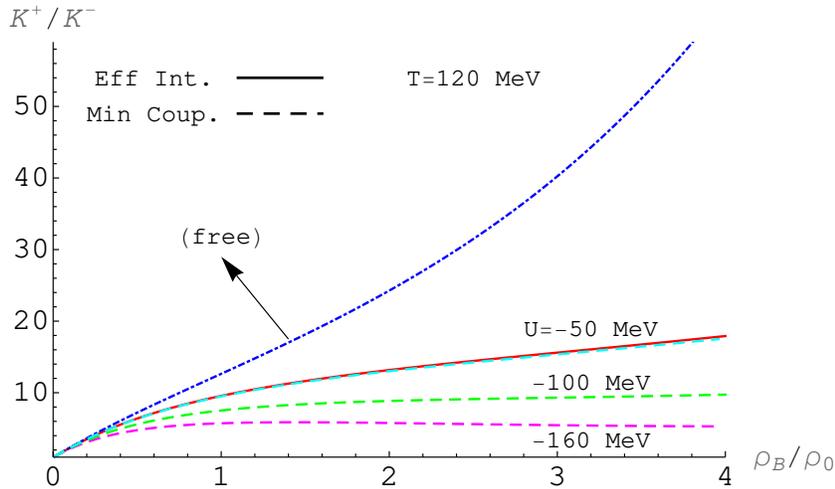,width=11cm}}
\vspace*{8pt}
\caption{Kaon to anti-kaon ratio as a function of baryon density at
a fixed temperature $T=120$ MeV. The solid lines correspond to the results obtained in the effective relativistic mean field model, the dashed lines correspond to different values of anti-kaon optical potential and the dot-dashed lines to a non-interacting free kaon gas.}
\end{figure}
\begin{figure}[ht]
\centerline{\psfig{file=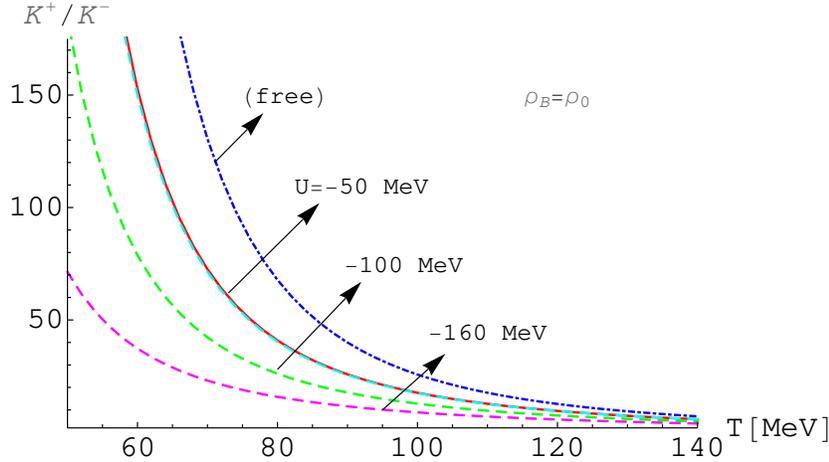,width=11cm}}
\vspace*{8pt}
\caption{The same of Fig. 2 as a function of temperature at a baryon density fixed to the nuclear saturation density $\rho_0$.}
\end{figure}

In this context it is interesting to observe that the results obtained in the minimal coupling scheme for a moderate potential depth ($U_{K^-}=-50$ MeV), as suggested by recent self-consistent calculations based on chiral Lagrangian \cite{lutz,lutz2} and coupled-channel G-matrix theory \cite{tolos}, are very close to those of the effective relativistic mean field model.
Contrariwise, in absence of kaon interaction (dot dashed line) and especially at higher baryon density, we observe a strong increase in the $K^+/K^-$ ratio. This effect is mainly due to the absence of an effective kaon mass and chemical potential.

In Fig. 2, we report the variation of the kaon to anti-kaon ratio as a function of the temperature at a fixed baryon density $\rho_B=\rho_0$, where $\rho_0$ is the nuclear saturation density. We observe again a good correspondence between the two approaches for a moderate value of the anti-kaon optical potential.  In presence of a free kaons gas, there is a strong enhancement in the $K^+/K^-$, due to the absence of the $\sigma$ and $\omega$ fields self-interaction. The contribution of $\rho$-meson field appears to be instead negligible for this choice of parameter set ($Z/A = 0.4$) and especially at higher temperature, where anti-kaons are abundantly produced and the $K^+/K^-$ rapidly decreases. Such a sensible difference in the kaon to anti-kaon ratio can be considered as a relevant feature for the determination of the real part of the anti-kaon optical potential, in the future relativistic heavy ion collision experiments at high baryon density.

In Fig. 3, we show the strangeness concentration $Y_S$ of kaons ($K^+$) and anti-kaons ($K^-$), hyperons ($Y$) and anti-hyperons ($\overline{Y}$) as a function of the baryon density. The total strangeness is fixed to zero. As expected, almost all the strangeness fraction is carried by kaons and hyperons, whereas $K^-$ and anti-hyperons play a marginal role, contributing only at low baryon density and high temperatures.
\begin{figure}[ht]
\centerline{\psfig{file=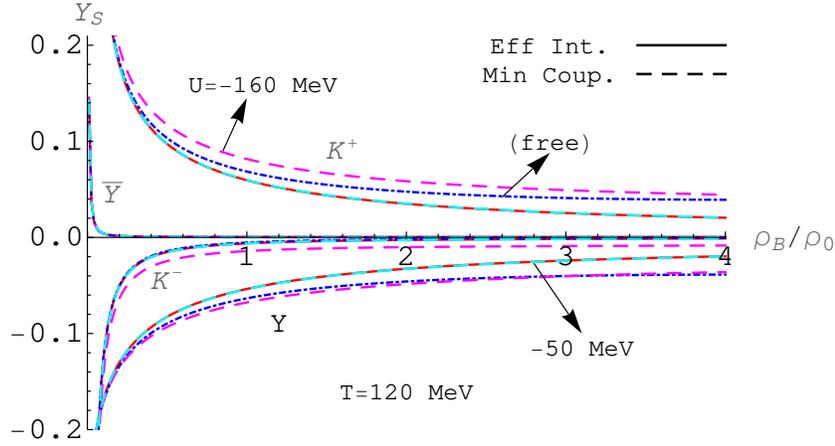,width=11cm}}
\vspace*{8pt}
\caption{Strangeness concentration as a function of baryon
density at a fixed temperature $T=120$ MeV. The solid lines correspond to the results obtained in the effective relativistic mean field model, the dashed lines correspond to different values of anti-kaon optical potential and the dot-dashed lines to a free kaon gas.}
\end{figure}
\begin{figure}[ht]
\centerline{\psfig{file=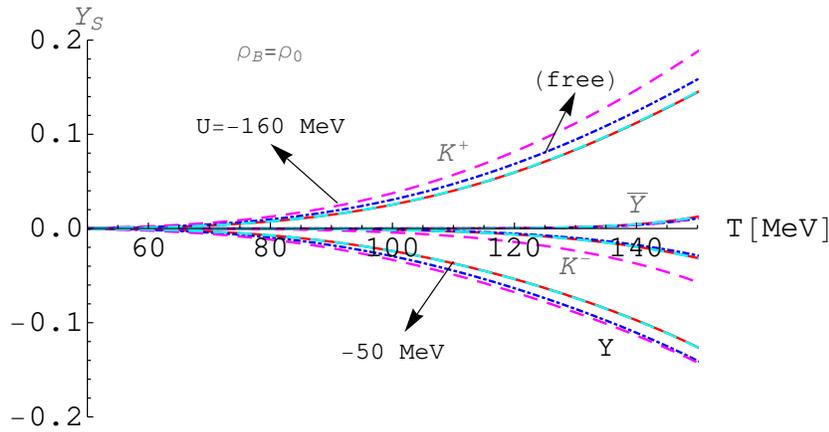,width=11cm}}
\vspace*{8pt}
\caption{The same of Fig. 3 as a function of temperature at a fixed baryon density $\rho_B=\rho_0$.}
\end{figure}

The results obtained through the relativistic mean field model, are in very good agreement with the minimal coupling scheme, when we set $U_{K^-}=-50 $ MeV. In absence of kaon interaction (free kaon gas, dot dashed lines), at low baryon density, the kaon strangeness concentration is close to that of the relativistic mean field model, whereas, at higher $\rho_B$ it converges to that of the effective minimal coupling scheme obtained for $U_{K^-}=-160$ MeV. Instead, anti-kaons strangeness fraction do not show an appreciable variation in absence of meson-field interaction.

In Fig. 4, we report the strangeness dependence of the system at the nuclear sa\-tu\-ration density $\rho_0$ for a wide range of temperatures. As we can see, the strangeness concentration is practically negligible below $T=60$ MeV. Strange particles start to be abundantly produced above $T=80$ MeV and the corresponding anti-particles are produced at higher temperatures. The free kaon gas case is also reported.
In this context, it is relevant to note that, for a choice of $U_{K^-}=-50$ MeV, the results obtained in the minimal coupling scheme are very close to those of the relativistic mean field one, because in this case we have an effective kaon mass close to the bare one. At the increasing of the anti-kaon optical potential, the effective kaon mass rapidly decreases, deviating from the behavior of the relativistic mean field model.
\begin{figure}[ht]
\centerline{\psfig{file=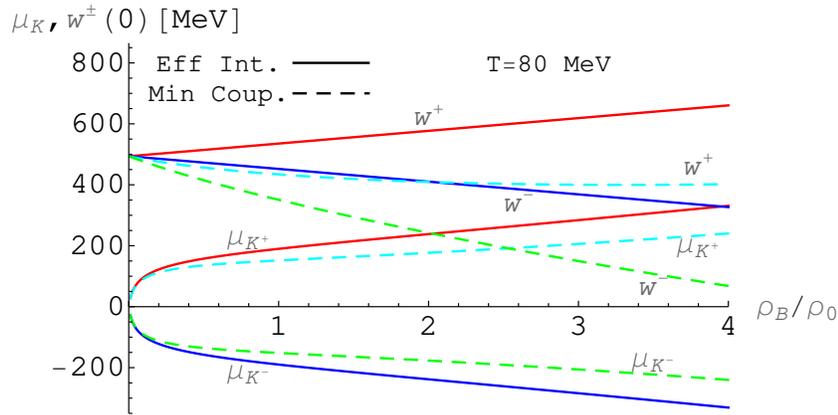,width=11cm}}
\vspace*{8pt}
\caption{Kaon and anti-kaon effective energy $\omega^{\pm}(p=0)$ and chemical potential as a function of baryon density at $T=80$ MeV in the effective relativistic mean field model (solid lines) and in the minimal coupling scheme (dashed lines) for $U_{K^-}=-160$ and $-50$ MeV.}
\end{figure}
\begin{figure}[ht]
\centerline{\psfig{file=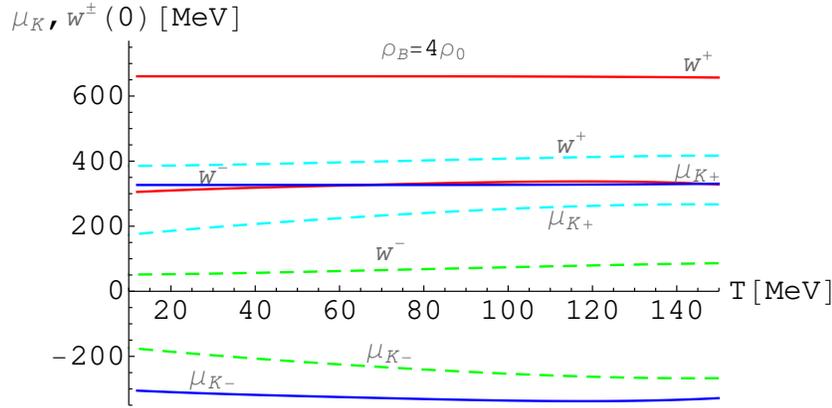,width=11cm}}
\vspace*{8pt}
\caption{The same of Fig. 5 as a function of the temperature at $\rho_B=4\rho_0$.}
\end{figure}

Before concluding, let us analyze the condition for the onset of the kaon condensation for a wide range of temperatures and baryon densities, in the two aforementioned models, considering $U_{K^-}=-50$ and $-160$ MeV. At this purpose, in Fig. 5, we report the threshold condition for the onset of the kaon and anti-kaon condensation ($\omega^{+}=\mu_{K^+}$ for $K^+$ and $\omega^{-}=\mu_{K^-}$ for $K^-$) at $p=0$ (s-wave condensation) as a function of baryon density and at the lower temperature at which strange particles start to be produced ($T=80$ MeV). We can see that the kaon/anti-kaon chemical potential $\mu_{K^{\pm}}$ is always lower than the corresponding kaon/anti-kaon threshold energy $\omega^{\pm}$, therefore the condition for the onset of the condensation is never reached. However, it is interesting to note that, for $U_{K^-}=-160$ MeV, $\mu_{K^+}$ approaches $\omega^+$, but they do not come close enough to allow the formation of the condensation. Note also that for $U_{K^-}=-50$ MeV the minimal coupling curves are perfectly overlapping to that of the effective relativistic one.
In absence of the kaon interaction, the effective kaon energy is obviously constant and equal to the kaon mass ($\omega^{\pm}=m_K$). The chemical potential is close to that of the minimal coupling scheme for $U_{K^-}=-160$ MeV, hence also in this case the threshold condition for the onset of the condensation is never reached.

Analogously, in Fig. 6, we report the variation of the kaon and anti-kaon effective energy and chemical potential for a very wide range of temperatures at the maximum baryon density explored in this work ($\rho_B=4\,\rho_0$). It is interesting to observe that both kaon and anti-kaon effective energies seem to be very little affected by variation of the temperature, contrariwise kaon and anti-kaon chemical potentials show a little increase at higher temperature (such a feature persists also by increasing the isospin asymmetry $Z/A$). This behavior seems to suggest that in relativistic heavy ion collision kaons can never reach the threshold condition for the onset of the condensation. Such a result appears to be in agreement with the predictions of very recent transport models \cite{hartnack}. The situation is of course different in systems like neutron stars, where the total strangeness is not conserved and the kaon condensation can take place \cite{zakout,pal,thorsson,banik}.

\section{Conclusion}
In this work we have analyzed the strangeness production at fixed isospin density ($Z/A=0.4$) and for a zero net strangeness, over a wide range of baryon densities and temperatures.
At this scope we have used two different approaches, an effective relativistic mean field model, where kaons are treated as a quasi-ideal Bose gas, but with the introduction of an effective chemical potential depending on the self-consistent interaction between baryons, and a minimal coupling scheme, where kaons are directly coupled with the meson fields ($\sigma$, $\omega$ and $\rho$). In this context, we have studied the influence of the kaon in-medium interaction in the
determination of the kaon to anti-kaon ratio and in the strangeness production, for different values of the anti-kaon optical potential ($U_{K^-}=-50$, $-100$ and $-160$ MeV). The results have been also compared with the case of a noninteracting kaon particles. In this analysis, we have observed a good correspondence between the two models in particular for moderate values of the optical potential ($U_{K^-}\simeq -50$ MeV), suggested by recent self-consistent calculation based on chiral Lagrangian and G-matrix theory \cite{tolos,lutz,lutz2}. In this condition, the kaon to anti-kaon ratio and the strangeness fraction become very close to those of the effective relativistic mean field model. Contrariwise, when we take in consideration a stronger potential depth (e.g. $U_{K^-}=-160$ MeV), the results appear to be different, mainly due to the strong variation in the kaon effective mass in the meson-exchange model.

We have also seen that, in absence of kaon in-medium interaction, the kaon to anti-kaon ratio rapidly diverges by the results obtained in the meson exchange and the relativistic mean field model, mainly due to the absence of the contribution of the $\sigma$ and $\omega$ field  into the kaon mass and chemical potential. The kaon strangeness concentration is also sensibly modified taking a behavior intermediate between that of the relativistic mean field model, at low baryon density and the minimal coupling scheme at $U_K=-160$ MeV, for higher values of $\rho_B$. Whereas anti-kaons strangeness fraction do not show an appreciable variation in absence of meson-field interaction.
The strong difference in the kaon to anti-kaon ratio could be considered as a relevant feature in the determination of the real part of the anti-kaon optical potential, especially in relativistic heavy ion collisions at high baryon density \cite{senger,arsene,bravina}.

Finally, we have also analyzed the possible onset of the kaon condensation in regime of density and temperature reachable in relativistic heavy ion collisions. We have found that the kaons chemical potential is always less then the corresponding kaon threshold energy $\omega^{\pm}$). This matter of fact seems to suggest, in agreement with the results obtained within modern transport codes \cite{hartnack}, that kaon condensation does not take place at any temperature and density in those systems in rapid evolution, like the relativistic heavy ion collision, where the zero net strangeness condition is conserved.

\end{document}